\documentclass{JHEP3}

\usepackage{epsfig}

\newcommand{\be}{\begin{equation}}
\newcommand{\ee}{\end{equation}}

\title{Chiral symmetry restoration at finite temperature in the planar limit.}
\author{R. Narayanan
\\Department of Physics, Florida International University, Miami,
FL 33199, USA\\E-mail: \email{rajamani.narayanan@fiu.edu}}
\author{ H. Neuberger
\\ Rutgers University, Department of Physics and Astronomy,
Piscataway, NJ 08855, USA\\E-mail: \email
{neuberg@physics.rutgers.edu} }

\abstract {We investigate numerically chiral symmetry restoration
at finite temperature in the planar limit in the deconfined phase,
both when it is stable and when the system is supercooled. We find chiral symmetry
restoration at $T_\chi = T_d$, where $T_d$ is the temperature of the deconfinement
transition in pure gauge theory and $T_\chi < T_d$ in the supercooled deconfined phase.
In the stable case the spectrum of the Dirac operator opens a gap in a discontinuous
manner and in the supercooled phase the gap seems to vanish continuously.}

\preprint{}

\begin{document}

\section{Introduction.}

At finite temperature QCD undergoes qualitative changes of great
physical interest. Although much is known, the complicated
strongly coupled aspects of the transition region are not under
full theoretical control.

At a large number of colors, in the  't Hooft limit, the system
simplifies somewhat~\cite{pis}. The major effects occur now in the
pure gauge sector, while the fermions only react to these effects,
without influencing them by back reaction (except, as explained later,
by aligning the pure gauge vacuum). The purpose of this
paper is to present numerical results on chiral symmetry
restoration at finite temperature in the planar limit. Our study
is at zero quark mass.
Preliminary results have been presented last year~\cite{lat05}.

At infinite $N_c$ the free energy is temperature ($T$) independent
at order $N_c^2$ for $T  < T_d$, where $T_d$ is the deconfinement
temperature. Chiral symmetry is spontaneously broken and the
condensate $\frac{1}{N_c} \langle\bar\psi\psi\rangle$ is nonzero
and temperature independent~\cite{nna,tc}. The chiral symmetry
breakdown is reflected by a condensation of eigenvalues of the
Euclidean Dirac operator near zero. This condensation emerges
naturally in a random matrix context. Because of the infinite
number of colors and the lack of relevance of the size of the
system due to large $N_c$ reduction, one can think of the
Euclidean Dirac operator ($D$) as a large random anti-hermitian matrix,
whose structure is restricted only by chiral symmetry.
\be
D=\pmatrix {0 & C\cr -C^\dagger & 0 } \ee In the spirit of
Wigner's approach to complex nuclei, one is lead to write down the
simplest probability distribution for the matrix $C$~\cite{sv},
whose linear dimension is proportional to $N_c$:
\be
P(C)\propto e^{-\kappa \dim (C) Tr C^\dagger C }\label{prob}
\ee
Chiral symmetry breaking is then an immediate result, giving it the
appearance of a generic phenomenon.

As the temperature is raised, at $T=T_d$, a first order deconfinement
transition occurs at all $N_c \ge 3$; $T_d$ has a finite large $N_c$ limit~\cite{tep2,joe}.
For high temperatures, $T >> T_d$, one expects chiral symmetry to
be restored and, consequently, the random matrix viewpoint that
worked for $T < T_d$ to become invalid. The simplest way in which
chiral symmetry can get restored is for the Euclidean Dirac
operator to open a gap at zero. One might have thought that this
effect can be incorporated into an extended random matrix
model~\cite{step}, but, a numerical investigation to be described
later on, indicates that the types of random matrix models one
would naturally guess will not work when chiral symmetry is
unbroken. For $T<T_d$ random matrix theory applies also at finite
$N_c$, but, without going to the planar limit the argument does
not extend to high temperatures ~\cite{urs}, where there is no energy regime
dominated by Goldstone particles~\cite{rmt}. We see that going to
infinite $N_c$ does not help in this respect.

\section{Large $N_c$ in the deconfined phase.}

At $T>T_d$, in the deconfined phase, the free energy of the gluons starts depending
on $T$. Feynman diagrams containing fermion loops are still suppressed by
one power of $\frac{1}{N_c}$, so long as the number of flavors is kept fixed, as we
do.

In the Euclidean path integral formulation, physical finite
temperature is reflected by the ``time'' direction being
compactified to a circle of radius $\frac{1}{T}$ and
bosons/fermions having periodic/antiperiodic boundary conditions
in the time direction.  For $T<T_c$  the boundary conditions are
irrelevant, since the preservation of the related $Z(N_c)\to U(1)$
global symmetry washes them out. When $T>T_d$, the trace of
parallel transport round the temporal circle (Polyakov loop)
acquires a fixed phase and breaks spontaneously the associated
$Z(N_c)$. Which phase is picked is arbitrary in the absence of
fermions, as all phases have the same gluonic energy. When
fermions are present, although in general their contribution to
the free energy is subleading, they fix the phase of the Polyakov
loop to one specific value because the fermions break the $Z(N_c)$
explicitly and align the vacuum. It is physically plausible, and
supported by our numerical work, that the phase is chosen to make
the Polyakov loop positive. This preserves CP, but also making the
Polyakov loop negative would have. Other than aligning the vacuum,
fermions have no impact on the distribution of the gluonic fields
at infinite $N_c$.

\section{Lattice setup.}

We work on a hypercubic lattice of shape $L_4 L^3$. The gauge
action is of single plaquette type.

\begin{eqnarray}
S=\frac{\beta}{4N_c}\sum_{x,\mu\ne\nu} Tr[ U_{\mu,\nu}(x)
+U_{\mu,\nu}^\dagger (x) ] \\
U_{\mu,\nu}(x)=U_\mu (x) U_\nu (x+\hat\mu) U_\mu^\dagger (x+\hat\nu)
U_\nu^\dagger (x)
\end{eqnarray}
We define $b=\frac{\beta}{2N^2_c}=
\frac{1}{g^2N_c}=\frac{1}{\lambda}$ and take
the large $N_c$ limit with $b$ held fixed. As usual, $b$ determines
the lattice spacing $a$ and $\lambda$ is the 't Hooft coupling.
The gauge fields are periodic. $x$ is a four component integer
vector labeling the site, and $\mu$ labels a direction;
a unit vector in the $\mu$ direction is denoted by $\hat\mu$. The link matrices $U_\mu(x)$
are in $SU(N_c)$.

There is a $Z^4(N_c)$ symmetry under which
\begin{equation}
U_\mu (x)\rightarrow e^{\frac{2\pi\imath k_\mu }{N_c}} U_\mu (x)
\end{equation}
for all $x$ with $x_\mu = c_\mu$. The integers $c_\mu$ are
fixed, and the integers $k_\mu$ label the elements of
the $\mu$-th $Z(N_c)$ group; $c_\mu =0,1,..,L_\mu-1$. Changing the $c_\mu$'s
amounts to a local gauge transformation. We have $L_1=L_2=L_3=L$.

The Polyakov loop matrix is denoted by $P_4 (x)$ and defined by:
\begin{equation}
P_4 (x)=U_4 (x) U_4 (x+\hat 4) U_\mu (x+2\cdot \hat 4)..U_\mu(x+(L_4-1)\cdot \hat 4)
\end{equation}
Under the $Z(N_c)$ factor associated with the time direction,
$P_4 (x)$ gets multiplied by a phase.
The gauge invariant content of  $P_4 (x)$ is its set of
eigenvalues (the spectrum)
$e^{\imath\theta^P_i},~i=1,2...,N_c$. The ordering is not gauge invariant,
and there is a constraint that $\det P_4 (x) =1$. Under the $Z(N_c)$
transformation, the set of eigenvalues is circularly shifted by a fixed
amount. The spectrum of $P_4 (x)$ and of $P_4 (x+j\cdot\hat 4)$ are
the same for all $j=0,1,2....,L_4-1$.

For a fixed $b$ in a certain range, if $L$ is large enough, the
global $Z^3(N_c)$ symmetry associated with the spatial directions
is unbroken. In practice, it is even possible to work in
metastable phases, as long as these $Z(N_c)$'s are maintained.

Depending on $L_4 \le L$ the $Z(N_c )$ in the time direction may
be broken or not. Alternatively, for $L_4 \le L$, one can break
the time-$Z(N_c )$ by increasing $b$ (for $L_4 = L$, we view as
time the one particular, but randomly selected, direction that
breaks its $Z(N_c)$). The breaking point is the deconfinement
transition. There is no dependence on $L$ on either side of this
transition so long as $b$ is in the range which preserves the
spatial $Z^3(N_c)$~\cite{knn}; this will always be the case.

In lattice units we have $a T_d = \frac{1}{L_c(b)}$, where~\cite{knn}
\begin{eqnarray}
b\rightarrow b_I \equiv b e(b)~~~~~~~e(b)=\frac{1}{N} 
\langle Tr U_{\mu,\nu} (x) \rangle \\
L_c(b)~\sim 0.260(15) \left ( \frac {11}{48\pi^2 b_I }\right )^{\frac{51}{121}} 
e^{\frac{24\pi^2 b_I}{11}}
\end{eqnarray}
In~\cite{knn} it was found that planar gauge theory on the torus 
can exist in five phases, $0$c,$1$c,..,$4$c; the 
deconfined phase is the 1c phase in this notation.

The fermion action is given by the overlap Dirac operator
which preserves chiral symmetry exactly. This choice makes it possible
to pose the question of spontaneous chiral symmetry breaking in a clean way.

The massless overlap
Dirac operator~\cite{overlap}, $D_o$, is defined by:
\begin{eqnarray}
&D_o = \frac{1+V}{2}\nonumber\\
&V^{-1}=V^\dagger=\gamma_5 V \gamma_5
={\rm sign}(H_w (M))\gamma_5
\end{eqnarray}
$H_w (M)$ is the Wilson Dirac operator at mass $M$, which we shall choose
as $M=-1.5$. $M$ should not be confused with a bare quark mass.
\be
H_w (M)= \gamma_5 \left [
M+4 -\sum_\mu \left ( \frac{1-\gamma_\mu}{2} T_\mu +\frac {1+\gamma_\mu}{2}
T_\mu^\dagger \right ) \right ]
\ee
The $T_\mu$ matrices are the lattice generators of parallel transport and
depend parametrically and analytically on the lattice links $U_\mu(x)$.

The internal fermion-line propagator,
$\frac{2}{1+V}$ is not needed at infinite $N_c$, since fermion loops are suppressed.
For fermion lines continuing
external fermion sources we are allowed to use a slightly different
quark propagator~\cite{ext} defined by:
\begin{equation}
\frac{1}{A} =\frac{1-V}{1+V}
\end{equation}
$A=-A^\dagger$ and anticommutes with $\gamma_5$. The spectrum of $A$
is unbounded, but is determined by the spectrum
of $V$ which is restricted to the unit circle. 
Up to a dimensionful unit,
$A$ should be thought of as a lattice realization of
the continuum massless Dirac operator, $D$:
\begin{equation}
2|M| A \leftrightarrow D=\gamma_\mu \partial_\mu + .....
\end{equation}

Our main observable will be the smallest eigenvalue of the
non-negative matrix $-A^2$, which is the discrete version of
$D^\dagger D$, where $D$ is the continuum, Euclidean, Dirac
operator in a fixed gauge background. The gauge background varies
according to the pure gauge action and we shall look at the
induced probability distribution of the smallest eigenvalues of
$-A^2$.

Let $\lambda_{1,2}$ be the two lowest eigenvalues of $\sqrt{-A^2}$:
The dimensionless gap as a function of
the dimensionless temperature $t$, $g(t)$, is defined as 
the average over gauge configurations of $\lambda_1(b,L,L_4) L_c(b)$.
The dimensionless temperature itself is defined as $L_c(b)/L_4$.

At infinite $N_c$, $g$ will vanish when the symmetry is
spontaneously broken. If we find that $g$ is nonzero, we know that
chiral symmetry is restored. This is true because the single way
chiral symmetry transformations can avoid inducing naive relations
among correlation functions of physical observables is by the
matrix $A^{-1}$ becoming singular with finite probability. If $g
>0$ almost surely, this cannot happen and chiral symmetry is
preserved.

As we vary $L_4$ or $b$, $g$ will change. If the change is
discontinuous to or from zero, the transition is of first order;
otherwise, it is continuous. We shall find that $g$ jumps
discontinuously from zero to a finite value exactly when the pure
gauge field undergoes the phase transition. Also, we shall see
some evidence that when we lower the temperature $T$ below $T_d$,
but stay in a supercooled deconfined phase, eventually, $g$ goes
to zero at some temperature $T_\chi < T_d$; this chiral symmetry
breaking transition, occurring as the temperature is lowered in the
supercooled deconfined phase, seems continuous within our
numerical resolution.

\section{Vacuum alignment.}

We wish first to check that indeed the vacuum of the gauge fields
is selected by aligning with the fermions in the manner discussed
earlier.

Numerically, we would like to show that the (positive) fermion
determinant is indeed favoring a positive Polyakov loop in the
deconfined phase. We do this as follows:

We let the lattice system pick an arbitrary phase for its Polyakov
loop in the deconfined phase. We now define the fermions with
twisted boundary conditions relative to this phase in the time
direction. That is, the fermions obey the boundary condition
$\psi(0)=-\psi(\frac{1}{T})e^{\imath(\theta-\phi)}$, where the
Polyakov loop has phase $e^{\imath \phi}$. We now intend to show
that the fermion determinant is maximal when $\theta=0$.

A complete computation of the determinant is too expensive and
might be an overkill. We accept the hypothesis that the
determinant is maximal when the gap $g$ is. After all, the
eigenvalues of the Dirac operator repel and a larger $g$ simply
would push all eigenvalues to slightly higher values, hence
increasing the determinant itself.

In summary we end up focusing on the gap $g$ as a function of the
angle $\theta$. As expected, we obtain a periodic function
of $\theta$ with a local maximum at $\theta=0$ and local minima
at $\theta=\pm \pi$, symmetric under $\theta \to -\theta$ and
monotonically decreasing from its value at $\theta=0$ to its value
at $\theta=\pi$.

\FIGURE[ht]{
\epsfig{file=gapbc.eps,  width=\textwidth}
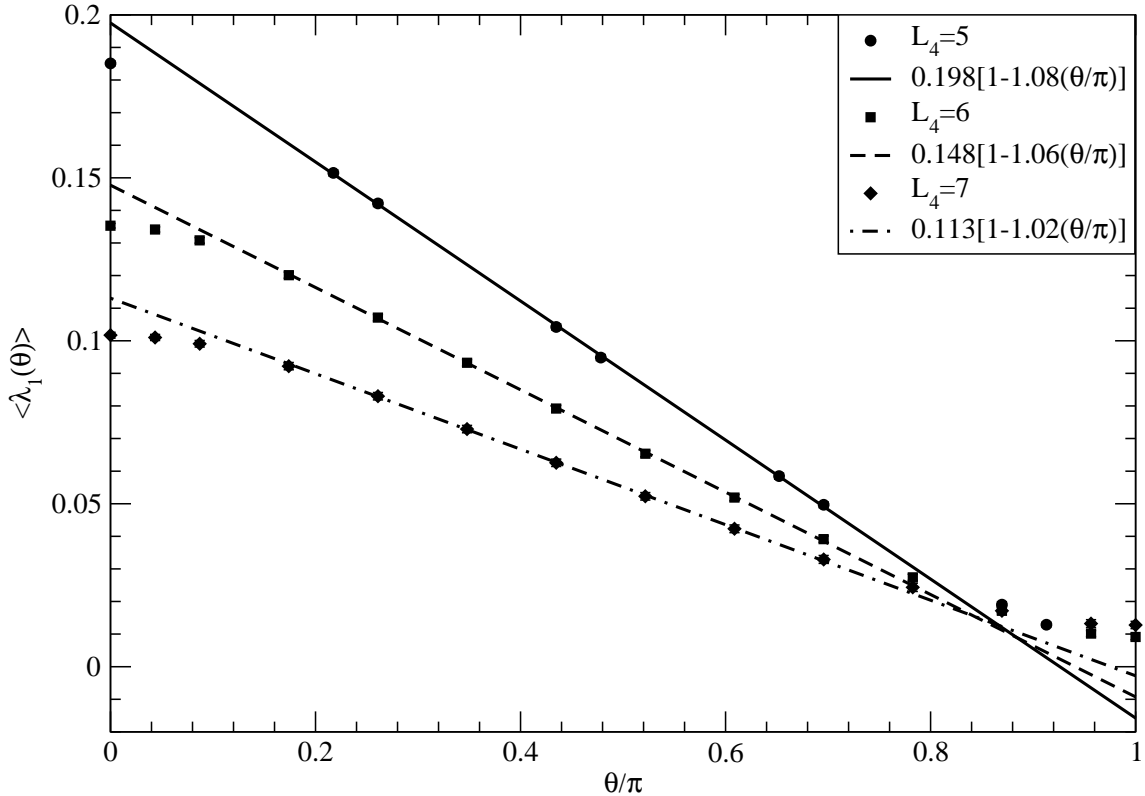
\caption{ The gap as a function of twist. }
\label{fig1}}

CP invariance implies $\frac{dg}{d\theta} =0$ at $\theta=0,\pi$;
figure~\ref{fig1} also shows that the region of maximum is
connected by an approximately linear segment to the regions of the
minima.

This linearity can be understood if one accepts a static
approximation, which is plausible for high enough temperatures. In
this static approximation the gauge field in the time direction is
taken as a constant, and the gauge fields in the space directions
are taken as time independent. In the continuum, the spectrum of
the operator $\gamma_4 D$ (where 4 labels the time direction) has
its spectrum linearly shifted by $\frac{T\theta}{\pi}$; assuming
now that this shift gets transmitted almost intact to the lower
bound of the operator $D^\dagger D$, we obtain the linearity of
the gap $g$ with a predicted slope.

Note however, that as the temperature is lowered towards $T_d$,
$T$ gets replaced by a smaller temperature, $T_{eff}<T$. Still,
the gap is maximal at $\theta=0$ and vanishes at $\theta=\pm\pi$,
and the linear portion is always present.

One might have expected to find the gap vanishing before $\theta$
reaches $\pm\pi$, for the following reason: Parallel transport
round the time direction results in a unitary matrix whose
spectrum has support on a connected arc on the unit circle,
centered at unity. Because of the finite width, one could have
imagined that in some color direction fermions are effectively
defined with a twisted boundary condition and the static
approximation would have predicted an earlier point where the gap
associated with that color direction vanishes. We do not see this
happening, indicating that there is no sense in which the
eigenvalues associated with the Polyakov loop can be used to coherently select
special orientations in color space for the fermions. All
that one can expect is an incoherent effect, in which the spread
in eigenvalues influences averages over all colors -- a point we
shall return to later.

In~\cite{step} the static approximation was used to motivate a
variation on the random matrix model describing the Dirac operator
in the confined phase. The Dirac operator is now split into blocks
labeled by the Matsubara integer $n$, and each block has a
structure
\be
D_n = \pmatrix { 0 & C + \imath [(2n+1) \pi - \theta]T\cr
-C^\dagger + \imath [(2n+1) \pi - \theta]T & 0} \label{step}\ee
with a common matrix $C$ distributed according to equation \ref{prob}.

In this picture the Polyakov loop is taken as a unit matrix with
one overall phase. If we generalized the above random model by
further splitting into blocks associated with each color we would
obtain a prediction that the $g(\theta)$ ought to vanish before
$\theta$ reaches $\pm \pi$, but this is somewhat implausible and
indeed does no happen as pointed out above.

\section{A possible implication of the spread in the eigenvalues
associated with the Polyakov loop.}

We present here a suggestion that the spread in eigenvalues
explains why one obtains an effective temperature $T_{eff} < T$
when one gets closer to $T_d$ from above.

Consider a free Dirac fermion in continuum with twisted boundary
condition $\theta$. Let $f$ be the contribution of this one
fermion to the free energy density in Euclidean space. Standard
manipulations~\cite{kapusta} produce
\be
f=4\left [ -\frac{\pi^4}{90}+\frac{\left ( 1-(\frac{\theta}{\pi})^2 \right )^2}{48}\right ]
T^4\ee

For $\theta=0$ we recover the well known expression. As $\theta$
increases toward $\pi$ the coefficient of $T^4$ decreases and
eventually even changes sign. Indeed, at $\theta=\pi$ we have
boundary conditions appropriate for a boson, but since we kept the
determinant at a positive power, we used for it Fermi statistics.
Had we used the right statistics, the determinant would have been
at a negative power and then the contribution to $f$ would have
had the normal sign.

We now speculate that the fact that parallel transport round the
compact direction is best described with a phase drawn from a
distribution symmetric about zero but not of zero width,
effectively implies an averaging over $\theta$ in the above
equation. In turn, this could be viewed as an effective reduction
of the temperature, if one insists on keeping the classical value
for the prefactor.

For large $T$, $T>>T_d$, the width of the distribution shrinks to
zero and the effect goes away. However, for $T$ close enough to
$T_d$, this indicates that $f\approx c(T)T^4$ where $c(T)$
decreases from its classical value at $T=\infty$ slowly. The
fermionic contribution to the free energy would therefore appear
almost noninteracting, however with a coefficient that is slightly
off.

Much more is needed to see if this speculation bears out.

\section{Supercooling.}

So long as we have confinement we ought to have spontaneous
symmetry breakdown at infinite $N_c$ for theoretical reasons \cite{cw}. We
also know from numerical work that the deconfinement transition in
planar QCD is of first order.

We now ask what happens to chiral symmetry as we supercool the
deconfined phase, to temperatures $T <T_d$. The gauge coupling
should increase, and eventually, chiral symmetry could break
again, this time without confinement. We address this question
numerically. Because of the need to extrapolate into the
metastable phase, the conclusion is quite tentative.

We see an indication that there is a chiral symmetry breaking
temperature $T_\chi < T_d$, where chiral symmetry is broken in the
supercooled deconfined phase. Our numerical finding is consistent
with a continuous transition at $T_\chi$.

Our numerical results are collected in table \ref{tab1}.
All the results are in the $1$c phase and we have used 
anti-periodic boundary conditions for fermions with respect
to the Polyakov loop in the broken direction. We studied
five different couplings, namely, $b=0.35$, $b=0.3525$, $b=0.355$,
$b=0.3575$ and $b=0.36$. We will use the
following central value for the critical sizes at these couplings:
\begin{eqnarray}
L_c(0.35)= 5.97;&&\ \ \  
L_c(0.3525)= 6.46;\ \ \ 
L_c(0.355)= 6.96;\cr
L_c(0.3575)= 7.47;&&\ \ \ 
L_c(0.36)= 8.01.
\end{eqnarray}
We performed a careful study at $b=0.3525$ and $L_4=L=6$
and convinced ourselves that the large $N$ limit is obtained
for values listed in the table. With the exception of the entry at 
$b=0.355$, $N=23$, $L=8$ and $L_4=7$, all entries are definitely 
in the stable $1$c phase. It is possible that the 
$b=0.355$, $N=23$, $L=8$ and $L_4=7$ entry is in the supercooled
$1$c phase.

\TABLE{
\caption{\label{tab1} Results for the lowest two eigenvalues
and their correlation $c$ as defined in (\ref{corr})
are listed for several values of $b$, $L$, $L_4$ and $N$
in the $1$c phase.}
\begin{tabular}{ccccccc}
$b$ & $N$ & $L$ & $L_4$ & $\langle \lambda_1 \rangle$ & $\langle \lambda_2 \rangle$ & $c$ \\
\hline
0.3500 & 43 & 6 & 2    & 0.7752(5)   & 0.7780(5)     & 0.94(2)  \\
0.3500 & 43 & 6 & 3    & 0.4199(7)   & 0.4249(6)     & 0.93(3)  \\
0.3500 & 43 & 6 & 4    & 0.2565(7)   & 0.2648(6)     & 0.68(8)  \\
0.3500 & 43 & 6 & 5    & 0.1591(10)  & 0.1693(8)     & 0.73(8)  \\
0.3525 & 43 & 6 & 2    & 0.7728(5)   & 0.7753(5)     & 0.95(2)  \\
0.3525 & 43 & 6 & 3    & 0.4189(6)   & 0.4233(6)     & 0.94(3)  \\
0.3525 & 43 & 6 & 4    & 0.2602(6)   & 0.2671(5)     & 0.79(5)  \\
0.3525 & 43 & 6 & 5    & 0.1674(11)  & 0.1771(10)    & 0.85(5)  \\
0.3525 & 31 & 6 & 6    & 0.1130(14)  & 0.1277(12)    & 0.71(7)  \\
0.3525 & 31 & 6 & 6    & 0.1069(21)  & 0.1219(18)    & 0.87(7)  \\
0.3525 & 37 & 6 & 6    & 0.1038(19)  & 0.1168(17)    & 0.89(3)  \\
0.3525 & 43 & 6 & 6    & 0.1091(15)  & 0.1223(10)    & 0.73(12) \\
0.3525 & 43 & 6 & 6    & 0.1049(16)  & 0.1179(12)    & 0.89(3)  \\
0.3525 & 47 & 6 & 6    & 0.1016(13)  & 0.1136(12)    & 0.82(4)  \\
0.3525 & 47 & 6 & 6    & 0.1030(13)  & 0.1162(12)    & 0.75(8)  \\
0.3525 & 53 & 6 & 6    & 0.1023(13)  & 0.1136(10)    & 0.81(6)  \\
0.3525 & 53 & 6 & 6    & 0.0992(14)  & 0.1116(12)    & 0.87(3)  \\
0.3550 & 43 & 6 & 6    & 0.1189(12)  & 0.1314(11)    & 0.82(8)  \\
0.3550 & 23 & 8 & 7    & 0.0623(17)  & 0.0767(13)    & 0.79(7)  \\
0.3575 & 43 & 6 & 6    & 0.1327(12)  & 0.1417(12)    & 0.89(3)  \\
0.3575 & 43 & 6 & 6    & 0.1325(11)  & 0.1423(10)    & 0.89(3)  \\
0.3575 & 23 & 8 & 7    & 0.0868(13)  & 0.0975(11)    & 0.86(5)  \\
0.3600 & 37 & 6 & 2    & 0.7610(6)   & 0.7636(5)     & 0.97(1)  \\
0.3600 & 37 & 6 & 3    & 0.4141(6)   & 0.4196(4)     & 0.82(5)  \\
0.3600 & 37 & 6 & 4    & 0.2654(8)   & 0.2725(7)     & 0.79(7)  \\
0.3600 & 37 & 6 & 5    & 0.1849(9)   & 0.1929(8)     & 0.87(4)  \\
0.3600 & 37 & 7 & 5    & 0.1832(6)   & 0.1891(5)     & 0.80(5)  \\
0.3600 & 37 & 7 & 5    & 0.1823(7)   & 0.1891(4)     & 0.76(8)  \\
0.3600 & 23 & 8 & 5    & 0.1837(7)   & 0.1904(6)     & 0.82(5)  \\
0.3600 & 23 & 8 & 6    & 0.1310(9)   & 0.1395(7)     & 0.75(6)  \\
0.3600 & 23 & 8 & 7    & 0.0963(10)  & 0.1064(8)     & 0.80(11) \\
0.3600 & 37 & 8 & 5    & 0.1807(5)   & 0.1861(4)     & 0.80(6)  \\
0.3600 & 37 & 6 & 6    & 0.1329(13)  & 0.1430(12)    & 0.91(3)  \\
0.3600 & 43 & 6 & 6    & 0.1353(10)  & 0.1445(9)     & 0.83(6)  \\
\hline
\end{tabular}
}

The data at the largest $N$ from table \ref{tab1} was used to obtain
the dimensionless gap, $g(t)$, as a function of the dimensionless 
temperature $t$. Figure \ref{fig2} shows all the data for $t < 2.8$.
There is
some spread of points, indicative of order $a^2$ 
lattice corrections, but the data
seems to condense toward a line to which we assign a physical meaning in the continuum.
We show a two parameter fit to
\be
g=1.76 \sqrt{t-0.93}
\ee
The square root behavior was imposed on the fit.

This indicates a continuous chiral restoring transition in the supercooled phase at $T\approx 0.93 T_d$.
We guess the transition ought to be continuous because the density of eigenvalues of the Dirac
operator at zero determines it, and so long as we stay in the deconfined phase (metastable regions
included) the dependence on the temperature of the gauge background has no a priori reason to change
discontinuously; one could view the temperature as entering only through the gauge background, but this
point might be debated.

\FIGURE[ht]{
\epsfig{file=lowests.eps,  width=\textwidth}
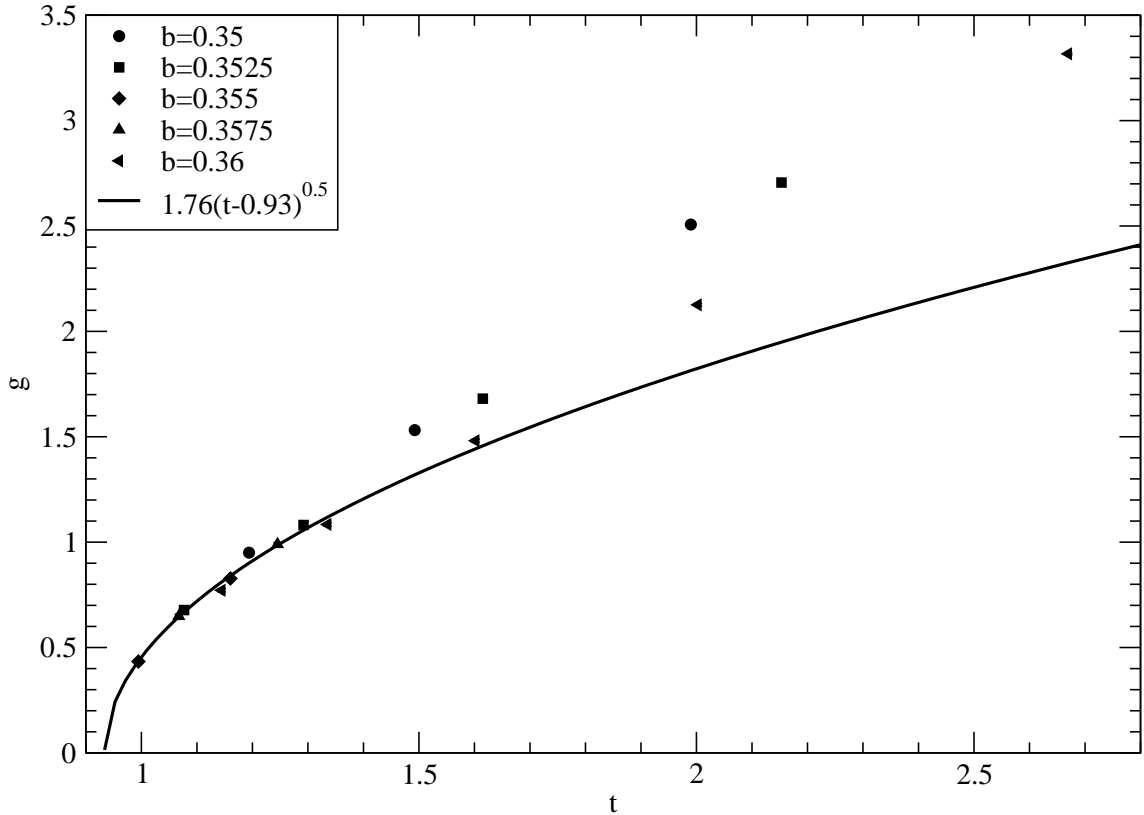
\caption{ The gap in units of $T_d$ as a function of temperature in units of $T_d$. }
\label{fig2}}

\section{No naive random matrix description.}

Intuitively, one could argue that large $N_c$ alone, without the
additional input from the viability of an effective chiral
Lagrangian description is enough to motivate a random matrix
description of spectral properties of the Dirac operator in planar
QCD. If this were true, one ought to be able to describe the
spectrum of the Dirac operator by some random matrix model even at
temperatures where chiral symmetry is restored.

We tried to see if this is possible by looking at the correlation
between the fluctuations of the lowest eigenvalues of the Dirac
operator. These fluctuations are about the mean, so independent of
the gap $g$, which is the main difference between the case where
chiral symmetry is broken and we know that random matrix theory
works and where we are when chiral symmetry is restored.

If indeed there is some random matrix model of the usual type
considered the correlation between such fluctuations is mainly
governed by level repulsion. We have two examples of such random
matrix models: one where chiral symmetry is broken and the other
where it is not, due to a large enough shift of $C$ (which occurs
for $T$ large enough) in the formula for $D$ in equation
\ref{step}.

We estimate by Monte Carlo methods the correlation, $c$
between the two lowest eigenvalues:
\be
c=\frac{ \langle (\lambda_1 - \langle \lambda_1 \rangle )
(\lambda_2 - \langle \lambda_2 \rangle )\rangle }{\sqrt{ \langle
(\lambda_1 - \langle \lambda_1 \rangle )^2 \rangle \langle
(\lambda_2 - \langle \lambda_2 \rangle )^2\rangle }} 
\label{corr}
\ee
In the two random matrix models mentioned before $c$ is between
$\frac{1}{2}$ (deep in the symmetric phase)
and $\frac{1}{3}$ (broken phase). $c$ is
bounded from above by 1 by a simple positivity argument. 
The numbers in table \ref{tab1} show
that $c$ is quite close to unity
deep in the symmetric phase
and and, although it drops down a little as
the temperature decreases, $c$ remains well above $0.5$
even close to $t=1$. This indicates a much stronger
correlation between the fluctuations of the lowest eigenvalues of
the Dirac operator than one would expect in any random matrix
model. The correlation we find is so strong as to imply an almost
rigid relationship between the two lowest eigenvalues.

If there exists a random matrix model that applies to the case
when chiral symmetry is restored, we are missing an essential
ingredient, which we speculate might have something to do with the
spectra of the eigenvalues associated with the Polyakov loops.

\section{Comparison to holographic models.}

Recently several papers discussing finite temperature features of
models that bear various degrees of semblance to QCD have
appeared~\cite{hol1,hol2,ofer}. These models are in the continuum,
but admit dual descriptions which allow control of the strong
coupling regime in the planar limit.

There are several models involving a set of branes which produce
the gauge fields and the dual gravitational background 
and ``probe'' branes, that have no effect on the 
background, which produce the quark fields. One finds
deconfining and chiral symmetry breaking transitions, all of first
order.

In particular, ~\cite{ofer} addresses the issue of a chiral
symmetry restoration transition in the supercooled deconfined
phase and finds it is of first order. It seems that a crucial
ingredient is the presence of an extra scale, beyond
$\Lambda_{QCD}$, which allows for a more complex phase diagram.
Nevertheless, it is noteworthy that some of these solvable
situations are members of continuously varying families, which
also contain ordinary QCD, albeit in a regime that is not under
control in the dual variables. 

Apart from the order of the transition in the supercooled phase,
which admittedly is only tenuously determined here, there is
general agreement between our findings here and the holographic
results.

\section{Conclusions}

Our main conclusion is unsurprising: at infinite $N_c$ the strong
first order deconfinement transition induces also immediate chiral
symmetry restoration.

At a more minor level we obtained some less expected results: If we
supercooled the deconfined phase chiral symmetry would still break
spontaneously and seemingly does so in a continuous transition.
Moreover, that transition would occur at a temperature which is
only seven percent below the deconfinement transition. It is well
known that such ``pseudo-transitions" are indicative of complex
dynamics even in the stable phase. We also saw that there is
something fundamentally different in the statistics of the
eigenvalues between the Dirac operator and typical random matrix
theory models. It would be worthwhile to understand why this is so
and what it means. In addition we were led to a heuristic
explanation for why the coefficient $c$ in the free gas law for
the free energy, $f=cT^4$, is weakly temperature dependent for
temperatures close to $T_d$ and smaller than the classical value.
The explanation viewed this as a consequence of the fact that the
eigenvalues associated with the Polyakov loop  are distributed
over an arc of a certain width, and only at infinite temperature
does the Polyakov loop become the trace of a matrix proportional
to the unit matrix.

The new tools of holographic duals provide a wealth of exactly
soluble strongly coupled theories where similar phenomena occur
and it would be useful to find observables and models where both
lattice and holographic methods apply simultaneously.

\acknowledgments

R. N. acknowledges support by the NSF under grant number
PHY-0300065 and from Jefferson Lab. The Thomas Jefferson National
Accelerator Facility (Jefferson Lab) is operated by the
Southeastern Universities Research Association (SURA) under DOE
contract DE-AC05-84ER40150. H. N. acknowledges partial
support by the DOE under grant number DE-FG02-01ER41165 at Rutgers
University and the hospitality of IFT, UAM/CSIC. 
We thank Ofer Aharony, Poul Damgaard and Jac Verbaarschot for useful exchanges.

\end{document}